# Ten computational challenges in human virome studies

Yifan, Wu[1], Yousong Peng[1, *]

[1]Bioinformatics Center, College of Biology, Hunan Provincial Key Laboratory of Medical Virology, Hunan University, Changsha, China

*Corresponding author. Email: pys2013@hnu.edu.cn

**Abstract**

In recent years, substantial advancements have been achieved in understanding the diversity of the human virome and its intricate roles in human health and diseases. Despite this progress, the field of human virome research remains nascent, primarily hindered by the absence of effective methods, particularly in the domain of computational tools. This perspective systematically outlines ten computational challenges spanning various phases of virome studies, ranging from virus identification, sequencing quality evaluation, genome assembly, annotation of viral taxonomy, genome and proteins, inference of biological properties, applications of the virome in disease diagnosis, interactions with other microbes, and associations with human diseases. The resolution of these challenges necessitates ongoing collaboration among computational scientists, virologists, and multidisciplinary experts. In essence, this perspective serves as a comprehensive guide for directing computational efforts in human virome studies, aiming to significantly propel the field forward.

**Introduction**

The human virome encompasses all viruses present in humans, including those infecting humans directly, viruses infecting bacteria, Archaea, and fungi, endogenous retroviruses, and viruses present as transients in food [1, 2]. It is estimated that there are approximately $10^{13}$ virus particles in humans [1]. In recent years, virome studies have identified a tremendous variety of viruses in humans, with the majority being phages, significantly expanding the diversity of the human virome [1, 3-5]. For instance, Shah et al. discovered 10,000 viral species from 248 virus family-level clades of the Caudoviricetes viral class, most of which were previously unknown, by sequencing of faecal viromes from 647 1-year-old children[4]. Nayfach et al. established the Metagenomic Gut Virus catalogue, which comprises 189,680 viral genomes from 11,810 public human stool metagenomes [3]. Lu et al. compiled 1137 animal and human viruses identified in human samples, including 68 human tissues, excreta, and body fluids, and constructed the Human Virus Database [5]. Viruses have been found in all major organs or tissues of humans. Liang's review provides a comprehensive description of the virome identified in different human organs or tissues [1]. However, there is still a lack of a reference virome in major human organs due to the large heterogeneity of the virome between individuals and a lack of effective and standard methods for accurate virome identification.

The virome has a significant impact on human health and diseases [1, 2, 6, 7]. Viruses can directly infect humans and cause acute or chronic diseases such as the flu, COVID-19, AIDS, and hepatitis. Long COVID suggests that viral infections may have a long-term influence on human health [8]. A recent study reported that infections with multiple viruses, such as influenza viruses and herpesviruses in early life, may increase the risk of neurodegenerative diseases such as Alzheimer's Disease (AD) and Parkinson's Disease (PD) [9]. Viruses can also influence human health by altering the dynamics of bacterial populations in the gut or other tissues [1, 2, 6, 7]. Dysbiosis of the microbiome, including the virome, has been frequently implicated in multiple gastrointestinal diseases, such as Inflammatory Bowel Disease (IBD) [6]. Although many associations between viruses and human diseases have been revealed, the influence of viruses on human health may still be overlooked. Moreover, the mechanisms underlying these associations are far from clear.

Human virome studies are still in the early stage, despite significant progress in recent years. The National Institutes of Health initiated the Human Virome Program in 2022, aiming to characterize the "healthy" human virome, overcome technological roadblocks, and define the virome's role in human health and disease [10]. One year later, the National Natural Science Foundation of China also initiated the Human Virome Project, aiming to develop novel methods for human virome studies and explore the composition and function of the virome in the respiratory tract of healthy individuals [11]. Obviously, the technologies and methods used in virome studies are a focal point of both funding programs. In fact, both experimental methods, from sampling to sequencing, and computational methods for analyzing virome sequencing data are still in an early stage of development compared to microbiome research [2, 7]. It is urgent to develop novel methods or improve current methods for human virome studies. This perspective illustrates the challenges in computational methods in human virome studies that need urgent resolution (Figure 1). Most of these challenges belong to the area of computational viromics, defined as using computational methods to address problems in viromics studies (see Lu's review for a comprehensive introduction to the area [12]). They call for more computational efforts from multiple disciplines to advance the field of the human virome.

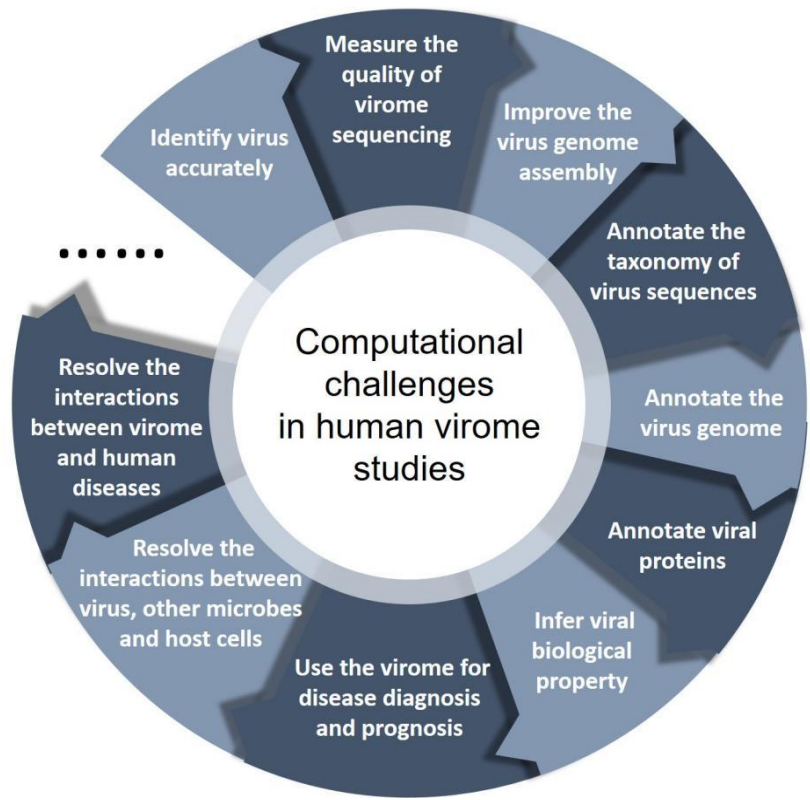

**Figure 1**. Ten computational challenges in human virome studies. Please see brief introductions about these challenges in texts.

## 1 How to identify virus sequences accurately?

The first step in the data analysis of virome studies is virus sequence identification. Unless viral-like particles (VLPs) are enriched before sequencing, less than 5% of sequencing reads typically belong to viruses in either metagenomic or metatranscriptomic sequencing [2]. Numerous methods have been developed to identify virus sequences, broadly grouped into homology-based and machine learning (ML)-based methods.

Homology-based methods are primarily employed for the identification of eukaryotic viruses, most of which have been well studied. The workflow of this method can be summarized as follows [13]: after removing reads from hosts, reads are assembled into contigs, or are directly aligned to a library of reference virus proteins or nucleic sequences using BLAST, leading to candidate virus sequences. Subsequently, for most studies, they are further aligned to the NCBI NT or NR database to eliminate false positives. Additionally, a stringent strategy should be applied to remove viruses with potential contamination, endogenous retroviruses, and viruses with low abundances. However, there is currently no unified workflow for this method of virus identification. No consensus has been reached on aspects such as the reference library for virus sequences (protein or nucleic sequences), the cutoff for determining candidate virus sequences, whether to remove false positives and contamination, etc., leading to inconsistency in virus identifications. For instance, Guo et al. identified 38 human-associated viruses from ten systemic lupus erythematosus (SLE) patients without removing any false positives or contamination [14], while Wu et al. identified only 10 human viruses from more than 600 SLE patients with a strict strategy [13]. Benchmark studies should be conducted to determine an optimal and unified workflow for virus identification.

ML-based methods for virus sequence identification are mainly used for the identification of prokaryotic viruses, which exhibit significant genetic diversity. Over a dozen such methods have been developed, with ten state-of-the-art methods benchmarked in Wu's study [15]. However, accurately identifying prokaryotic viruses remains challenging due to their extensive genetic diversity. Lu et al. estimated that there are at least 8.23e+08 viral Operational Taxonomic Units (vOTUs) on Earth, while less than 3% of viral genetic diversity has been uncovered so far [16]. ML-based methods only utilize known viruses, representing a very small proportion of the virus genetic space. Consequently, capturing the full spectrum of virus genetic diversity with current methods is challenging. High false positive rates have been observed for these methods, and there is inconsistency between prediction results from different tools. Therefore, it is recommended to use consensus results from multiple tools and then validate identified virus sequences using homology-based methods.

## 2 How to measure the quality of virome sequencing?

Various strategies are employed throughout the processes from sample collection to sequencing, and these strategies significantly impact the detection of viral sequences.

Here, we outline some common strategies (for a comprehensive review on these strategies, please refer to Mirzaei's review [17]): i) Enrichment of VLPs: the decision of whether to enrich VLPs and the method chosen for this enrichment have substantial consequences. Enriching VLPs can significantly enhance the detection of viral sequences but may introduce biases towards certain types of viruses. ii) Choice of sequencing approach: the decision to use metagenomic or metatranscriptomic sequencing introduces bias toward DNA or RNA viruses, respectively. iii) Sequencing platform: the selection between next-generation sequencing (NGS) and third-generation sequencing (TGS) platforms can lead to biases in the determination of virus abundances.

An essential question arises as to whether the virus reads obtained through sequencing accurately reflect the real diversity and abundance of the virome in the sample. Do these reads exhibit biases towards specific types of viruses, such as DNA or RNA viruses, enveloped or non-enveloped viruses, or viruses of particular sizes or shapes? Following the pre-processing of raw sequencing reads and the identification of virus reads, it becomes crucial to assess the quality of the viral reads. Unfortunately, methods for such quality measures are currently lacking. To address this, several indices can be provided to reflect the bias of virus reads. These may include the ratio of DNA to RNA viruses, the overall diversity of viruses, and other relevant metrics. Additionally, it is imperative to establish a reference virome in major organs or tissues through meta-analysis of existing virome sequencing data, such as the endeavors of the Metagenomic Gut Virus catalogue [3] and Human Virus Database [5]. Building a comprehensive reference virome will contribute to a better understanding of the biases in virus reads and aid in the accurate interpretation of virome data.

## 3 How to improve the virus genome assembly?

The assembly of virus genomes is an essential step for further studies of the virome. Unfortunately, accurately assembling virus genomes from short-read sequencing data remains challenging due to the presence of strong strain heterogeneity and the low abundance of viral populations. Presently, two main types of methods for virus genome assembly are employed [18]: reference-based assembly methods, such as MetaCompass and VirGenA, and *de novo* assembly methods, such as Haploflow. Reference-based methods assemble genomes by utilizing a known genome as a guide. While these methods are generally more accurate than *de novo* methods, they may not be suitable for viruses lacking high-quality reference genomes, particularly for novel viruses. A potential solution involves developing methods to assemble virus genomes by using the genomes of other viruses belonging to the same genus or even the same family as templates. In contrast, *de novo* methods can be applied to assemble genomes for all viruses, including novel ones, although they typically require deep sequencing depth and may not assemble complete genomes.

A promising approach to enhance virus genome assembly involves combining both types of methods. The integration of multiple assembly tools has been demonstrated in the state-of-the-art tool VIGA, developed for eukaryotic virus identification and genome assembly from NGS data [18]. Additionally, virome studies based on TGS can facilitate virus genome assembly since TGS generates long reads [19]. Combining the strengths of both NGS and TGS has the potential to further improve virus genome assembly.

**4 How to annotate the taxonomy of virus sequences?**

There are currently two main taxonomy systems for viruses: the International Committee on Taxonomy of Viruses (ICTV) and the Baltimore classification system. These systems classify viruses primarily based on biological features such as genome type, life cycle, virus shape, and more. However, for newly identified viruses lacking isolation and culture, no such biological features are available for classification. Recently, the ICTV has started to incorporate genomic sequences for classification. Nevertheless, unlike the 16S RNA in bacteria and the 18S RNA in eukaryotes, which serve as universal marker genes for taxonomy, there is a lack of a universal marker gene in viruses. Consequently, the ICTV can only capture a small proportion of viruses, especially prokaryotic viruses. In recent years, despite the identification of a significant number of viruses, more than half of newly identified viruses cannot be classified using current taxonomy systems [3, 4].

Various computational efforts have been undertaken to enhance virus classification [20]. For instance, vConTACT has established a virus classification system based on the gene content of viruses [20]. However, it relies on long contigs, lacks interpretability, and is unable to classify novel viruses that exhibit little homology to known viruses. Additional efforts are essential to construct a virus classification system that is universal, user-friendly, and interpretable. The development of such a system is crucial to accommodating the growing number of newly discovered viruses and ensuring a comprehensive understanding of viral diversity across different environments.

**5 How to annotate the virus genome sequences?**

The annotation of viral genomes is a prerequisite for further characterization of viruses. Due to their small genome size, viruses tend to maximize the utilization of genome sequences for encoding. Compared to their hosts, viruses exhibit few introns, and different genes often share sequences. In addition to proteins, viruses encode a diverse array of RNA products, including small RNAs, circular RNAs, long non-coding RNAs, and more [21, 22]. This diversity poses a significant challenge for virus genome annotations.

Although computational tools such as Prodigal and GeneMarkS are commonly used in virus genome annotations and generally perform well, most of these tools were not specifically developed for viruses [12]. Developing novel tools for virus genome annotation or enhancing existing tools by considering virus-specific features could substantially improve the accuracy of virus genome annotations. Additionally, combining multiple annotation tools may further enhance the precision of virus genome annotations. For instance, Wu et al. achieved high-accuracy annotation of the SARS-CoV-2 genome by combining GeneMarkS, ORFinder, and BLAST [23]. Furthermore, the emerging field of metaproteomics [24], defined as the large-scale identification and quantification of proteins from microbial communities, holds promise for improving gene identification in virome studies. While predicting protein-coding genes is relatively well-established, predicting RNA products from the genome remains challenging. The RNAs generated in metatranscriptomic sequencing offer valuable insights and may assist in the identification of RNA products in virus genomes, particularly in studies using TGS-based metatranscriptomic sequencing.

**6 How to annotate viral proteins?**

The inference of the structure and functions of viral proteins serves as the foundation for further studies on viruses. Current methods for annotating viral proteins primarily rely on alignment-based sequence homology analysis using tools like BLAST and HMMER [25]. However, many viral proteins exhibit little or no sequence homology with known proteins, posing a significant challenge for protein annotations. Advancements in Artificial Intelligence (AI) offer a potential solution to this problem. For instance, AlphaFold has successfully predicted three-dimensional structures for over 200 million proteins, demonstrating the power of AI in structural prediction [26]. While the AlphaFold Protein Structure Database currently lacks viral proteins, our study has shown that AlphaFold can predict confident structures for more than 40% of human virus proteins (data not shown). Additionally, a recent study by researchers from Meta introduced a novel method that directly infers full atomic-level protein structures from primary sequences using a large language model [27]. They applied this approach to predict structures for more than 600 million metagenomic protein sequences. In terms of function inference, another recent study utilized a protein language model to predict the functional categories of viral proteins with encouraging performances and they further expanded the annotated fraction of ocean virome viral protein sequences by 37% based on the model [25]. These developments indicate that the latest AI models have the potential to uncover the "secrets" of protein structure and functions from primary sequences, offering a promising framework for the annotation of viral proteins.

**7 How to infer the biological properties of viruses from genomic sequences?**

Inferring the biological properties of viruses, such as host specificity, pathogenicity, virulence, life cycle, and shape, is crucial for comprehending viruses and aligns with the requirements of the Minimum Information about an Uncultivated Virus Genome (MIUViG) standards [17]. Unfortunately, the majority of viruses have been identified at the sequence level, with only a few being successfully isolated and cultured. Consequently, over 95% of viruses in public databases are categorized as uncultivated, making it impractical to determine their biological properties through experimental means. Computational methods have become indispensable in inferring these properties, playing a key role in understanding the characteristics of viruses. For example, numerous computational methods, including both homology-based and ML-based approaches, have been developed for predicting virus hosts in both eukaryotic and prokaryotic viruses [17, 28]; the tissue tropism of human viruses have been predicted based on codon usage [29]. Despite notable achievements in these areas, accurately predicting viral biological properties remains computationally challenging for several reasons:

i) Limited training data: data for training ML methods to predict biological properties are often scarce.

ii) Complex determinants: many biological properties, such as host and virulence, are determined by multiple genes in both the virus and their hosts. The mechanisms underlying these properties remain unclear for most viruses.

iii) Diversity: viruses exhibit significant diversity, making it challenging, if not impossible, to develop a universal method applicable to all viruses.

iv) Rapid mutation: viruses mutate rapidly, leading to dynamic changes in biological properties. This rapid mutation rate hinders the development of effective prediction methods.

## 8 How to use the virome for disease diagnosis and prognosis?

Viruses have been recognized for their influence on human diseases, either by directly infecting humans or by shaping the bacterial community. Consequently, the virome can serve as a reflection of the healthy status of humans. Notably, certain viruses, like torquetenovirus, have been identified as indicators reflecting the state of the immune system [30]. Utilizing the virome for disease diagnosis is a logical extension of this understanding. In a manner similar to metagenomics studies, virus taxa are employed as features in ML-based disease diagnosis. For instance, in studies by Guo et al., ML models were constructed to classify SLE patients and healthy individuals based on virus taxa [14]. Similarly, Zhang et al. developed survival analysis regression models for prognostic predictions in several cancers, leveraging information from virus taxa [31]. However, the virome is characterized by high dynamism and heterogeneity. Most virus taxa at the species, genus, and family levels are observed in only a small proportion of samples, making accurate disease diagnosis or survival prediction challenging based solely on virus taxa. A potential solution involves considering more prevalent features, such as high-level taxa (phylum, class, and order), virus protein families, gene functions, virus hosts, and macro-level features of the virome such as the alpha-diversity.

**9 How to resolve interactions between virus, other microbes and human cells?**

Viruses depend on hosts for their survival, engaging in intricate interactions with other microbes, including bacteria and fungi, as well as human cells. These interactions collectively shape micro-ecosystems or micro-environments that exert a profound influence on human health and diseases such as cancer, respiratory diseases, and gastrointestinal diseases. While extensive research has explored the interactions between viruses and their hosts, particularly phages and their bacterial hosts, there is a notable scarcity of studies investigating the intricate web of interactions involving viruses, other microbes, and human cells [32]. Compounding this challenge is the lack of effective methods for studying such multifaceted interactions. Network-based methods offer a promising avenue for delving into the complexities of these interactions [33]. Co-occurrence networks, for instance, between viruses and/or bacteria, serve as valuable tools for understanding the intricate dynamics at the population level. Constructing co-occurrence networks that encompass multiple types of organisms may provide a more accurate representation of micro-environment population dynamics. Moreover, at the gene level, a co-occurrence or co-expression network has the potential to unveil the mechanisms behind complex interactions among various organisms.

**10 How to resolve the interaction between virome and human diseases?**

Viruses play pivotal roles in human diseases, and the mechanisms underlying the association between viruses and human diseases are intricate. These associations can be broadly categorized into three types:

i) Direct causation: human viruses that directly cause human diseases, such as the influenza virus, SARS-CoV-2, HBV, and HIV. These viruses have been extensively studied and can induce obvious diseases in humans.

ii) Indirect causation: some human viruses, like herpesviruses, may remain dormant for extended periods without causing diseases. However, they can resurge and lead to fatal diseases when the human immune system is compromised. Additionally, certain viruses can have long-term effects on human health, as exemplified by long COVID resulting from SARS-CoV-2 infection [8]. Associations between several neurodegenerative diseases like AD and PD and early-life infections with viruses like Herpes Simplex Virus and influenza viruses have also been reported [9]. To infer such associations, both epidemiological studies and molecular-level interactions between viruses and diseases can be employed. For example, Teulière et al. found significant associations between virus infections and human aging by analysis of the human and viral protein interactomes [34].

iii) Disruption of microbial balance: prokaryotic viruses can cause human diseases by disturbing the balance of bacterial populations. Dysbiosis of the microbiome, attributed to phages, has been linked to gastrointestinal diseases like IBD [2, 7]. Associations may be established between specific virus taxa and diseases, but interpretability is limited as phages do not directly infect humans. Establishing associations between the characteristics of the virome population and diseases, particularly in the context of phage-induced dysbiosis, is more reasonable. However, it is crucial to note that these associations are correlations rather than causal relationships. Functional studies in animal models are essential to elucidate causal relationships, a topic explored in detail in the review by Chaffringeon et al. [35], emphasizing the use of animal models to understand the role of the virome in determining human health and disease.

**Conclusions**

Despite significant progress in the field of the human virome in recent years, it remains in its infancy, and substantial challenges persist. A primary impediment is the lack of effective methods, particularly in the realm of computational tools. This perspective highlights ten computational challenges across various stages of virome studies, encompassing virus identification, sequencing quality evaluation, genome assembly, annotation of viral taxonomy, genome and proteins, inference of biological properties, applications of virome in disease diagnosis, interactions with other microbes, and associations with human diseases. It's important to note that this list may not be exhaustive, and as the field continues to advance, new computational challenges may emerge. Addressing these challenges necessitates a concerted effort to propel the field of human virome forward. Continuous collaboration between computational scientists, virologists, and other experts is essential to drive innovation and develop comprehensive solutions to the computational challenges that currently impede progress in the study of the human virome.

**Acknowledgements**

The authors thank members in Peng's lab for helpful discussions on the manuscript.

**Funding**

This work was supported by National Natural Science Foundation of China (32170651 & 32370700).

**Availability of data and materials**

Not applicable.

**Competing interests**

The authors declare that they have no competing interests.